\documentclass[
 Reprint,PRX,twocolumn,
superscriptaddress,
 amsmath,amssymb,
 aps,
]{revtex4-2}

\usepackage{graphicx}
\usepackage{dcolumn}
\usepackage{bm}
\usepackage{xr}
\usepackage[dvipsnames]{xcolor}
\usepackage{color}
\usepackage[normalem]{ulem}

\begin{document}

\preprint{APS/123-QED}

\title{Breaking of scale invariance in a strongly dipolar 2D Bose gas}

\author{Haoting Zhen}
\thanks{These authors contributed equally.}
\affiliation{Department of Physics, The Hong Kong University of Science and Technology,
Clear Water Bay, Kowloon, Hong Kong}%

\author{Yifei He}
\thanks{These authors contributed equally.}
\affiliation{Department of Physics, The Hong Kong University of Science and Technology,
Clear Water Bay, Kowloon, Hong Kong}%

\author{Sampriti Saha}
\affiliation{Institut f\"ur Theoretische Physik, ETH Z\"urich, Wolfgang-Pauli-Str. 27 Z\"urich, Switzerland.}%

\author{Mithilesh K. Parit}
\affiliation{Department of Physics, The Hong Kong University of Science and Technology,
Clear Water Bay, Kowloon, Hong Kong}%

\author{Mingchen Huang}
\affiliation{Department of Physics, The Hong Kong University of Science and Technology,
Clear Water Bay, Kowloon, Hong Kong}%

\author{Nicol\`o Defenu}
\affiliation{Institut f\"ur Theoretische Physik, ETH Z\"urich, Wolfgang-Pauli-Str. 27 Z\"urich, Switzerland.}%

\author{Gyu-Boong Jo}
\altaffiliation{email: gbjo@rice.edu}
\affiliation{Department of Physics and Astronomy, Rice University, Houston, TX, USA}%
\affiliation{Smalley-Curl Institute, Rice University, Houston, TX, USA}%
\affiliation{Department of Physics, The Hong Kong University 
of Science and Technology,
Clear Water Bay, Kowloon, Hong Kong}%


\date{\today}
\begin{abstract}
Two-dimensional (2D) dipolar atomic gases present unique opportunities for exploring novel quantum phases due to their anisotropic and long-range interactions. However, the behavior of strongly dipolar Bose gases in 2D remains unclear, especially when dipoles are tilted. Here, we demonstrate the creation and characterization of strongly dipolar 2D condensates in a quasi-2D harmonic trap with tunable dipole orientation. By investigating scale invariance properties through breathing collective mode measurements, we observe significant breaking of scale invariance when dipoles are tilted in-plane indicating the dominance of the nonlocal dipole-dipole interactions (DDIs) in this regime. Interestingly, the breaking of the scale invariant dynamics is accompanied by an increase in quantum fluctuations, as shown by comparison with mean-field and beyond mean-field theoretical studies. Our experiments also reveal that at critical tilt angles around 70°, stripe-type density modulations emerge, suggesting the presence of a roton spectrum in 2D, while the system still shows hydrodynamic nature with the phase-locking breathing behavior. This observation elucidates the many-body effect induced by DDIs in 2D, thus marking a crucial step toward realizing 2D supersolids and other exotic quantum phases. 
\end{abstract}

\maketitle

Two-dimensional (2D) bosons with dipole-dipole interactions (DDI) involve both isotropic short-range and anisotropic long-range interactions. In contrast to 2D Bose gases with contact interactions only - where the interaction-induced Berezinskii-Kosterlitz-Thouless (BKT) mechanism \cite{hohenberg1967existence,berezinskii1972destruction,kosterlitz1972long} reveals superfluidity with quasi-long-range order \cite{hadzibabic2006berezinskii,desbuquois2012superfluid,murthy2015observation} - the DDI stabilizes various crystalline structures, potentially leading to the emergence of supersolid order. While interesting scenarios can occur when dipoles are tilted, a systematic understanding of strongly dipolar 2D systems remains elusive, especially how crystalline structure emerges with superfluidity. In this work, we take advantage of the scale invariance, a fundamental property resulting from length-scale-free short-range interaction in 2D, and examine how strong DDI breaks such invariant property
~\cite{hung2011observation,yefsah2011exploring,desbuquois2014determination,murthy2015observation} and manifests itself in the 2D superfluid~\cite{vasiliev2014universality}.

Recently, a single-layer quasi-2D dipolar superfluid with tunable dipolar angles was realized with moderate DDIs $\epsilon_{dd}\sim0.5$~\cite{he2025exploring} (where $\epsilon_{dd}=a_{dd}/a_s$ represents the ratio of dipolar to contact scattering lengths). While mild anisotropic effects of DDIs manifest in density fluctuations for tilted dipoles, an effective contact-like treatment of DDIs sufficiently describes the BKT superfluid transition, and scale-invariant behavior is observed from equations of state.  Nonetheless, DDIs are expected to produce more pronounced effects in the strongly dipolar regime with $\epsilon_{dd}\gtrsim1$ in 2D, where intriguing phenomena such as anisotropic superfluidity~\cite{ticknor2011anisotropic}, anisotropic vortex cores~\cite{yi2006vortex}, anisotropic vortex-vortex interactions~\cite{mulkerin2013anisotropic}, and 2D stable bright solitons~\cite{pedri2005two,nath2009phonon} are predicted.

Another open question is the possible existence of supersolid in 2D~\cite{recati2023supersolidity} when density modulation induced by roton instability~\cite{santos2003roton} coexist with 2D superfluidity. As a phase with crystalline structure and global phase coherence, supersolid has been observed in elongated or oblate dipolar gases in 3D~\cite{tanzi2019observation,chomaz2019long,bottcher2019transient, norcia2021two,bland2022two}. In 2D, complex supersolid phases are expected to be found in a strongly dipolar gas with perpendicular dipoles at zero temperature~\cite{zhang2019supersolidity,zhang2021phases,roccuzzo2022supersolid,ripley2023two}, though the required density is far higher than current experimental capability. Recently, several works found that a strongly dipolar 2D gas with tilted dipoles can potentially host a stripe-like supersolid with much lower density~\cite{mishra2016dipolar,bombin2017dipolar,staudinger2023striped,aleksandrova2024density,SanchezBaena2025Tilted,lima2025supersolid}, which coincides with the solid-like density modulations observed in the strongly dipolar gas trapped in a quasi-2D pancake trap~\cite{wenzel2017striped}. In the same setup, however, no experimental evidence of phase coherence could be found. This is in agreement with the theoretical picture in Refs.~\cite{cinti2019absence,cinti2020comment}, where the thermodynamic stability of 2D supersolid has been questioned.
\begin{figure*}
    \includegraphics[width=0.75\textwidth]{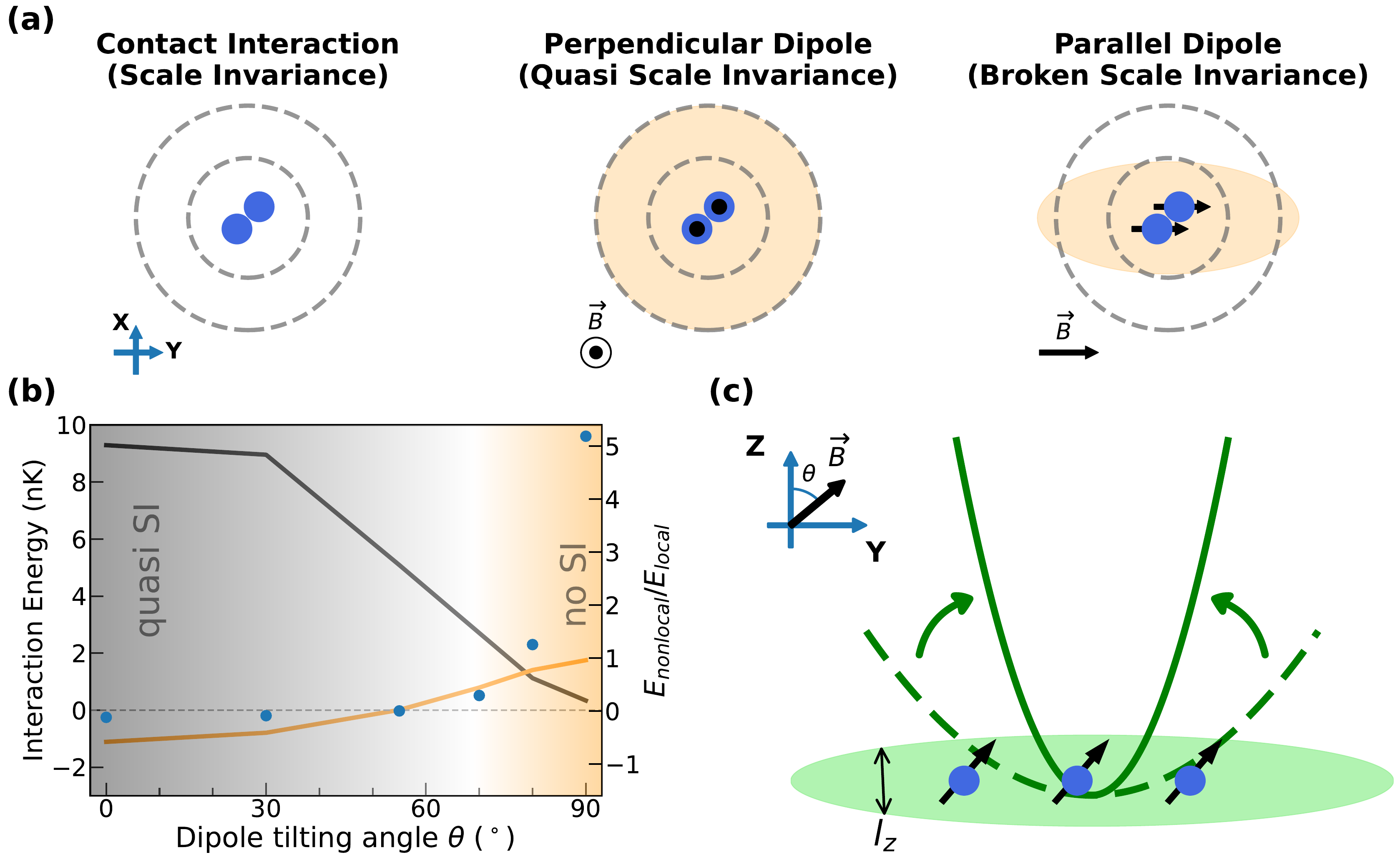}
    \caption{\footnotesize \textbf{Scale-invariance theory and the breathing mode in a quasi-2D dipolar gas.} \textbf{a}, Breaking of scale invariance (SI) in a quasi-2D dipolar gas. At the classical field level, isotropic contact interactions preserve SI due to their short-range nature. When dipoles are aligned perpendicular to the 2D plane, DDIs (shown in yellow) are largely isotropic and short-range, resulting a quasi-scale-invariance (quasi-SI). In contrast, SI is broken when dipoles lie parallel to the 2D plane, where local interactions are suppressed and the anisotropic nonlocal part of the DDIs become significant. \textbf{b}, Energy contributions from local contact interactions and nonlocal DDIs as a function of dipole angle $\theta$ with $\epsilon_{dd}=0.98$. The black and orange solid lines represent the local and nonlocal interaction energies, respectively, computed from a simulated ground-state profile (see text). Dots indicate the ratio $E_{\text{nonlocal}}/E_{\text{local}}$. Shaded regions and the dashed line serve as visual guides. \textbf{c}, Experimental setup. A sheet beam tightly confines atoms along the z-axis, with an oscillator length $l_z\sim 0.24\mu m$. An external magnetic field $\vec{B}$, rotating in the $y$-$z$ plane, controls the dipole orientation. A quench of the radial trap excites the breathing mode in the $x$-$y$ plane.} 
    \label{fig1}
\end{figure*}

Here, we produce strongly dipolar 2D condensates in a quasi-2D harmonic trap with variable dipole orientation. We examine the scale invariance properties with DDIs by measuring the breathing mode. A strong breaking of scale invariance is observed when dipoles are tilted in plane, indicating the dominance of the nonlocal interaction involved with $1/r^3$ scaling of dipole interactions. By increasing $\epsilon_{dd}$ over 1, we find the breathing mode frequency displays a larger up-shift and observe a stripe-type density modulation with tilt angle at $70^{\circ}$, which suggest the softening of the roton excitations in the 2D spectrum. Lastly, we study the phase-locking behavior of the breathing mode, validating the nature of our strongly dipolar 2D condensates.

The collective modes we use in this work are fundamental properties for understanding a many-body state, which is associated with the low-lying excitation of quantum gases. In particular, the breathing mode is a sensitive probe of collective properties such as interaction~\cite{jin1996collective,yan2024collective} and correlation~\cite{kao2021topological}, it has been extensively used to characterize different physical processes such as dimensional crossover~\cite{merloti2013breakdown,de2015collective,holten2018anomalous,peppler2018quantum} and phase transitions~\cite{chomaz2016quantum,ferrier2018scissors,tanzi2019supersolid,tanzi2021evidence,chisholm2024probing,huang2025probing}. In an elongated dipolar gas, the lowest-lying collective mode is used to manifest the simultaneous breaking of the phase invariance and the translational symmetry, which establishes the supersolid nature in the sample~\cite{tanzi2019supersolid}. Here, we utilize this powerful tool to explore the competition between local and nonlocal interactions in a strongly dipolar 2D superfluid, which provides valuable insights on the distinctive nature of different phases of matter existing in 2D dipolar gas.

\section*{\bf Non-local part of DDI and scale invariance} 
 For a 2D Bose gas with only contact interactions, the ratio between breathing mode frequency and trap frequency in 2D exhibits a universal value $\omega_B/\omega_t=2$, and the interaction dependence to the breathing mode frequency, which is caused only by quantum fluctuations, is negligible~\cite{pitaevskii1996dynamics,chevy2002transverse,vogt2012scale,saint2019dynamical}. This universal property stems from the $SO(2,1)$ symmetry, which arises from the reduction of the conformal invariance of the homogeneous gas by the parabolic trapping potential~\cite{pitaevskii1997breathin}. In Bose gases, typical interaction strengths are not strong enough to observe this correction~\cite{hung2011observation}, but observations of quantum scale anomaly~\cite{olshanii2010example} (see Methods) have been reported in strongly-interacting 2D Fermi gases~\cite{holten2018anomalous,peppler2018quantum}.

\begin{figure*}
    \includegraphics[width=1.0\textwidth]{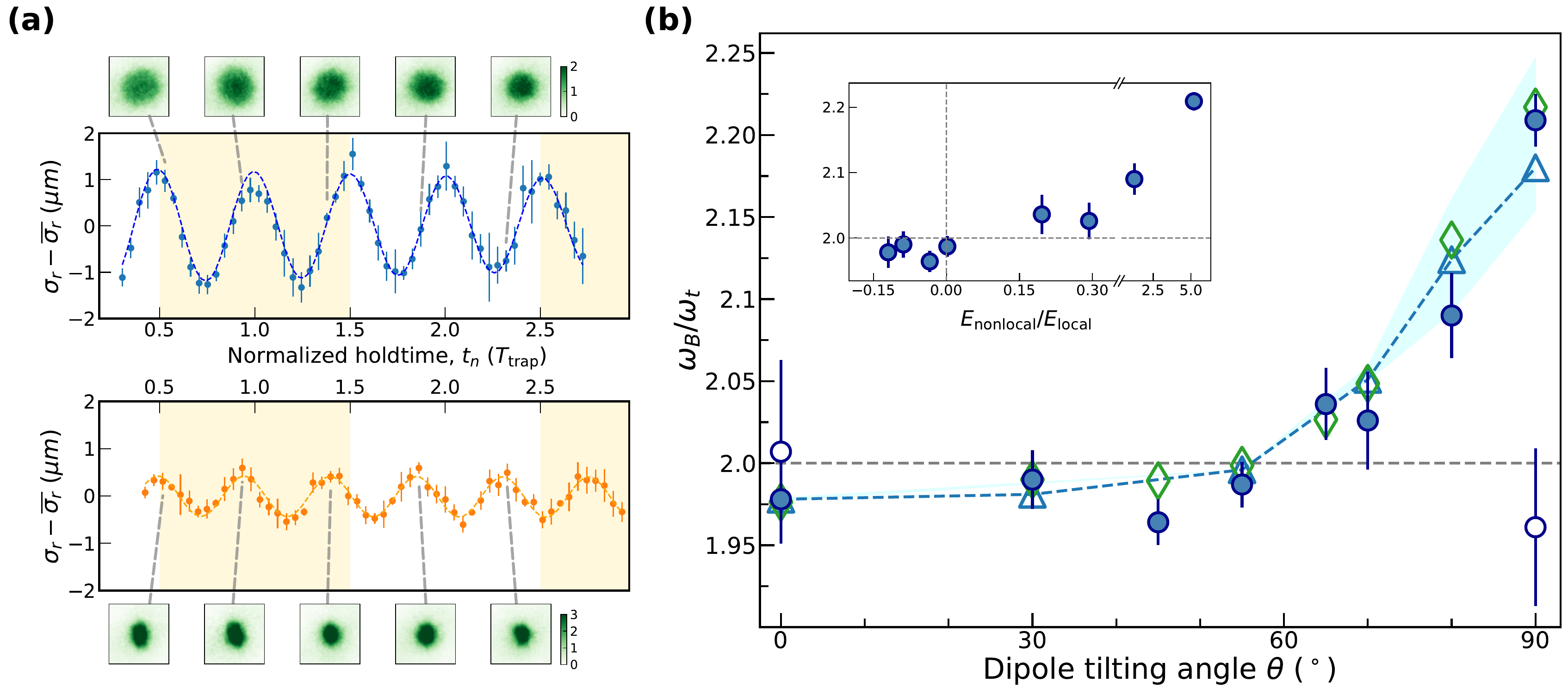}
    \caption{\footnotesize \textbf{Breaking of scale invariance in coherent dipolar samples with tilted dipole orientation $\theta$.} \textbf{a}, Example measurements of the breathing mode dynamics. Top and bottom panels show in-situ images for $\theta=0^{\circ}$ and $90^{\circ}$ samples, respectively, at $\epsilon_{dd}=0.98$, displayed as the optical density. Central panels depict oscillations in the average radial width $\sigma_r$, plotted relative to its mean value over all hold times, $\overline{\sigma_r}$. The holdtime $t_n$ is normalized by the trap period $T_{\text{trap}}$. Dashed lines are damped sinusoidal fits~\cite{supplementary}.
     The shades representing one trap period are guides to eyes. \textbf{b}, The breathing frequency of a quasi-2D dipolar gas with $\epsilon_{dd}=0.98$ and different dipole orientations. The blue (white) points correspond to the breathing mode frequencies of condensates (normal gas). As dipoles rotate into the 2D plane, the breathing mode frequency of the condensates gradually deviates from the $SO(2,1)$ symmetry value $\omega_B=2\omega_t$, represented by the grey dashed line. The mean-field GPE simulation is shown as triangles, and the blue dashed lines are direct connection of the symbols. The results of the MCTDHB simulation with $M=2$ are shown as green open diamonds. Incorporating quantum fluctuations does not significantly alter the GPE predictions, except at large $\theta$, where quantum effects shift the frequency upward, in agreement with experimental observations. The blue shade represents the calibration uncertainty on $a_s$~\cite{patscheider2022determination}. The inset compares the breathing mode frequency to the relative strength of nonlocal DDIs. Error bars represent the 95$\%$ confidence interval.} 
    \label{tiltfig}
\end{figure*}

For quasi-2D dipolar gases where atoms occupy the ground state of the axial harmonic confinement, interactions can be considered as a combination of a contact-like part and a nonlocal part~\cite{cai2010mean}. The former $E_{\text{local}}\propto \widetilde{g}_{\text{eff}}\int n(r)^2dr^2$ describes the contribution of contact interaction and the local part of DDI, where the effective interaction strength $\widetilde{g}_{\text{eff}}\propto a_{\text{eff}}/l_{z}=(a_s+a_{dd}(3\cos^2{\theta}-1))/l_{z}$ is dimensionless. Thus, the value $\widetilde{g}_{\text{eff}}$ is determined by both the 3D s-wave scattering length $a_{s}$ and the dipolar length $a_{dd}=\mu_0\mu_m^2m/12\pi\hbar^2$ in units of the transverse oscillator length $l_{z}=\sqrt{\hbar/m\omega_{z}}$, which confines the atoms to the 2D plane. $\theta$ is the angle between the dipoles and the $z$ axis. The nonlocal part of DDIs can be expressed by~\cite{ticknor2011anisotropic},
\begin{align}
    E_{\text{nonlocal}}&\propto \widetilde{g}_{dd}\int d\textbf{r}^2 n(\textbf{r})\int \frac{d\textbf{k}^2}{(2\pi)^2} e^{i\textbf{k}\cdot\textbf{r}}F_d\left(\frac{kl_z}{\sqrt{2}}\right)n(\textbf{k}), \\
    F_d(x)&=x e^{x^2}\text{erfc}(x)(-\cos^2{\theta}+\sin^2{\theta}\cos^2{\phi_k}),
    \label{fd}
\end{align}
where $\widetilde{g}_{dd}\propto a_{dd}/l_{z}$ is the effective 2D coupling strength of DDIs, $\text{erfc}$ is the complementary error function, and $\phi_k$ is the angle between $k$ and the projection of dipole orientation on the 2D plane. The kernel $F_d$ captures the long-range nature of DDIs in quasi-2D, as it is explicitly dependent on the magnitude of the momentum $\textbf{k}$. While $E_{\text{local}}$ preserves scale invariance, $E_{\text{nonlocal}}$ does not, as it can be confirmed by inspecting the isotropic case. For $\varphi_{k}\approx 0$, the function $F_d(x)$ becomes flat at large momenta $k\gg l_{z}$ while behaving linearly for $k\lesssim l_{z}$. Thus, the two momentum region transform differently with $k\to k/\lambda $, and  scale invariance is broken. 

In a quasi-2D dipolar gas, $E_{\text{nonlocal}}$ typically becomes small compared to $E_{\text{local}}$ due to the relatively large effective local interaction strength $\widetilde{g}_{\text{eff}}$\,\cite{fischer2006stability}. In this case, the contribution of nonlocal DDIs can be neglected and the scale invariance is largely preserved, which is observed in 2D Townes solitons~\cite{chen2021observation} and 2D dipolar superfluid~\cite{he2025exploring}. While stronger nonlocal DDIs can be found in systems with larger dipole moments such as polar molecules~\cite{bigagli2024observation}, it can produce significant effects in weaker dipolar systems if local interactions are strongly suppressed. Such suppression can be achieved by either decreasing $a_s$ via Feshbach resonance or tilting dipoles into the 2D plane, as reflected in the form of $\widetilde{g}_{\text{eff}}\propto (a_s+a_{dd}(3\cos^2{\theta}-1))$. A schematic of this effect is shown in Fig.~\ref{fig1}a. The scale invariance is largely preserved in a 2D dipolar gas with out-of-plane dipoles where local interactions dominate. Only with dipoles tilted in-plane, the anisotropic nonlocal DDIs significantly alter the scaling dynamics of the system. This effect can be visualized by the computation of the interaction energies from both local and nonlocal DDIs based on a simulated ground-state density profile~\cite{supplementary}. As shown in Fig.~\ref{fig1}b, as $\theta$ increases, nonlocal interactions gradually dominate, driving the system into the scale invariance-breaking regime.


\section*{\bf Breaking of scale invariance with tilted dipoles}

To investigate scale invariance in presence of DDIs, we perform our experiments employing a spin-polarized gas of $^{166}$Er, which is loaded into a quasi-2D trap and probe breathing mode oscillation frequency $\omega_B$ in the strongly dipolar regime with $\epsilon_{dd}\gtrsim 1$ (see Methods for more details). Here, the angle $\theta$ serves as a control parameter to investigate how gas dynamics evolves as scale invariance is progressively broken. To compare $\omega_B$ with the standard value of $2\omega_t$, we determine the radial trap frequency using dipole modes simultaneously excited by the same quench. By performing a sinusoidal fit  to the oscillation of the cloud center, we obtain the trap frequencies $\omega_{x}$ and $\omega_{y}$ along the two trap principal axes~\cite{supplementary}. The mean trap frequency is calculated as $\omega_t=\sqrt{\omega_x \omega_y}$.

Fig.~\ref{tiltfig}a shows an example measurement of the breathing mode. The top and bottom panels display the in-situ density profiles of the sample with $\theta=0^{\circ}$ and $\theta=90^{\circ}$, respectively, imaged at different hold times. The sample with $\theta=90^{\circ}$ appears smaller and denser due to the reduction in repulsive local interactions~\cite{he2025exploring}. At the five selected hold times, the $\theta=90^{\circ}$ sample is maximally expanded, whereas the $\theta=0^{\circ}$ sample contracts from its maximum width to the minimum. This contrast is reflected in the oscillations of $\sigma_r$ for both angles, as shown in the central panels. After normalizing the hold time $t$ by the trap period $2\pi/\omega_t$, we observe that the sample with $\theta=90^{\circ}$ oscillates significantly faster than that with $\theta=0^{\circ}$, indicating a breaking of scale invariance.

Unlike the case of local interactions, where such violations typically arise from quantum fluctuations~\cite{olshanii2010example}, here the breakdown occurs already at the semi-classical level due to the $\propto r^{-3}$ scaling of dipolar interactions. To systematically explore the impact of DDIs, we measure the breathing mode frequency across a range of dipole tilt angles, $\theta = 0^{\circ}, 30^{\circ}, 45^{\circ}, 55^{\circ}, 65^{\circ}, 70^{\circ}, 80^{\circ}$, and $90^{\circ}$, while keeping the dipolar strength fixed at $\epsilon_{dd} = 0.98$. As shown in Fig.~\ref{tiltfig}b, the ratio $\omega_B/\omega_t$ remains close to 2 when the dipoles are predominantly oriented out-of-plane. In this regime, local interactions dominate over nonlocal contributions, maintaining quasi-scale invariance with $E_{\text{interaction}} \approx E_{\text{local}}$.
As the dipoles are tilted further into the plane, the nonlocal component of the DDIs becomes increasingly significant, leading to a marked upshift in $\omega_B/\omega_t$. Here, the contact interaction and the local part of the DDIs nearly cancel, resulting in an effective local interaction strength $\widetilde{g}_{\text{eff}}$ approaching zero. Consequently, the residual nonlocal interactions govern the system's behavior.
The most pronounced deviation occurs at $\theta = 90^{\circ}$, where $\omega_B/\omega_t$ exceeds the standard value of 2 by approximately 10\%, signaling a strong violation of scale invariance. 

We also measure the breathing mode frequency of two samples in the non-condensate regime, with dipoles oriented perpendicular to and within the 2D plane. These samples are prepared with a lower atom number $N\approx 1.9\times10^4$, and a higher temperature $T\approx 1.6\hbar\omega_z$. As shown in Fig.~\ref{tiltfig}b, $\omega_B/\omega_t$ recovers the standard value of 2 within experimental resolution. No significant up-shift is observed with $\theta=90^{\circ}$, indicating that quasi-scale invariance persists in the strongly dipolar 2D thermal gas, independent of dipole orientation. This result already points to the deep connection between the breaking of scale invariance and the emergence of strong quantum fluctuations. In the thermal regime, quantum correlations are weak, rendering the contribution of nonlocal interactions also negligible~\cite{thermal}, see the inset Fig.~\ref{tiltfig}b. This measurement also demonstrates that the frequency of collective oscillation could serve as a probe for identifying the superfluid phase in a strongly dipolar 2D gas within the scale invariance-breaking regime.

The connection between quantum coherence in the 2D gas and the violation of $SO(2,1)$ symmetry highlights a subtle interplay between the long-range tails of the dipolar interaction—which explicitly break the $SO(2,1)$ invariance—and the quantum fluctuations inherent to the system. It is crucial to disentangle the effects of scale invariance breaking that emerge at the semi-classical level, from those—if present—that stem from the interplay between the dimensional nature of dipolar interactions and the many-body correlations within the gas.

To investigate this interplay, we adopt a multi-configurational time-dependent Hartree for bosons (MCTDHB) framework accounting quantum fluctuations~\cite{meyer1989,fischer2015condensate} as well as a time-dependent Gross-Pitaevskii equation~(GPE) solver at the semiclassical level~\cite{supplementary} to simulate the breathing mode dynamics of samples at various tilt angles $\theta$. In general, the semiclassical results (empty triangles) show good match with experimental measurements (full circles), except that a significant upward shift is observed in both experimental data and MCTDHB at $\theta=90^\circ$, as illustrated in Fig.~\ref{tiltfig}b. This upward shift from the semiclassical prediction, observed at $\theta= 90^{\circ}$, implies the effect of quantum fluctuation, which is accounted in the MCTDHB simulation with $M=2$ orbitals. A similar shift has been studied theoretically in 3D contact gas~\cite{pitaevskii1998elementary}. Interestingly, the most pronounced effect from quantum fluctuations coincides with the strongest breaking of scale-invariance at the semi-classical level.

In the inset of Fig.~\ref{tiltfig}b, we plot $\omega_B/\omega_t$ with respect to the relative strength of nonlocal and local interactions, showing a clear relationship between the $SO(2,1)$ symmetry violation and the dominance of nonlocal DDIs. Interestingly, from both the GPE simulation and the MCTDHB simulation, we find that a negative nonlocal interaction energy leads to a down-shift in $\omega_B/\omega_t$ relative to 2, while a positive nonlocal interaction energy causes an up-shift. Experimentally, it should be noticed that the down-shift below $55^\circ$ may arise from a combination of third-dimensional effects~\cite{merloti2013breakdown} and the negative nonlocal interaction energy.
\begin{figure*}
    \includegraphics[width=0.9\textwidth]{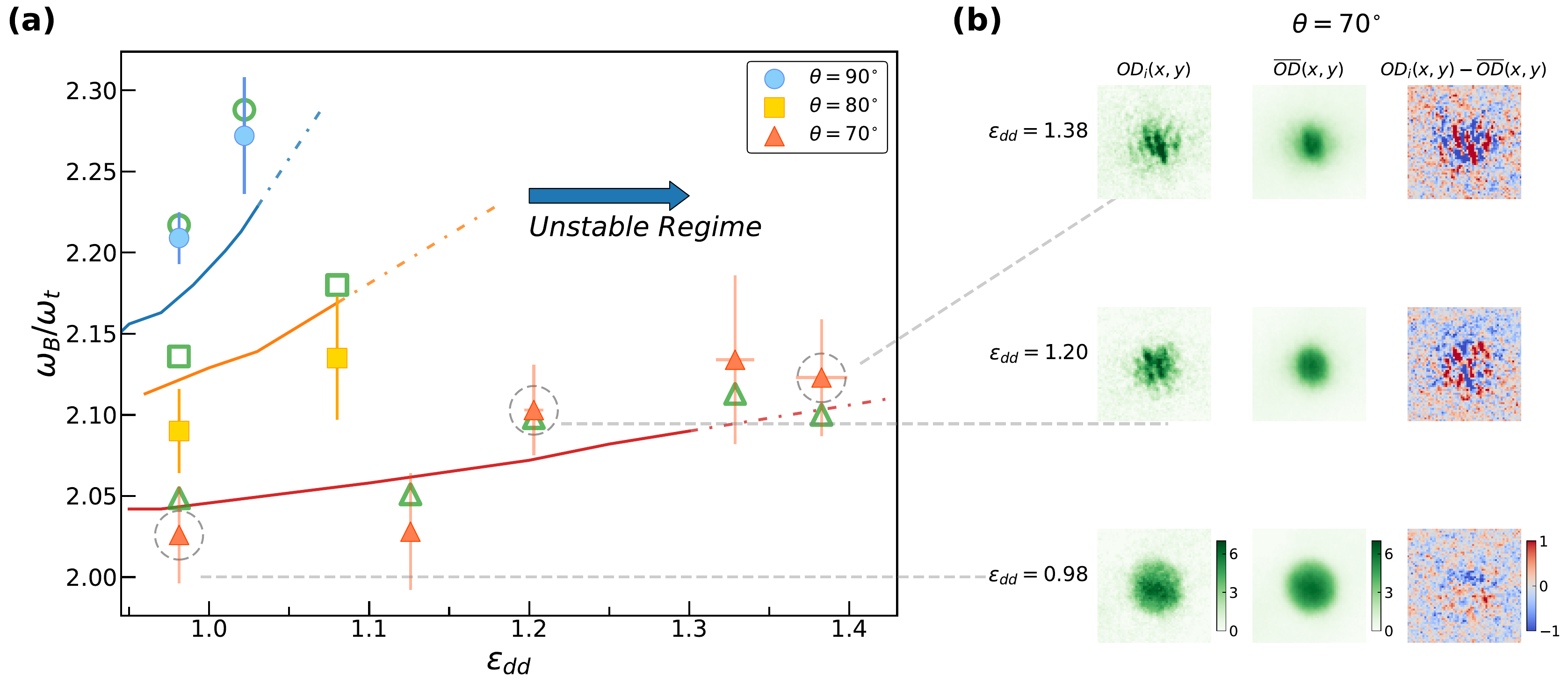}
    \caption{\footnotesize \textbf{The signatures of nonlocal DDIs in the strongly dipolar regime.} \textbf{a}, The breathing mode frequency of samples in the strongly dipolar regime with highly tilted dipoles. With larger $\epsilon_{dd}$, $\omega_B$ deviates more from the standard value. While the mean-field GPE simulation, plotted as solid lines (and the dash-dot line as extension in the mean-field unstable regime), captures the trend of this strong scaling anomaly, the MCTDHB simulation, shown as green labels, matches the measurements better by including the effect of quantum fluctuations. Error bars represent the 95$\%$ confidence interval. 
    \textbf{b}, The in-situ profiles of samples with $\theta=70^{\circ}$ and $\epsilon_{dd}=0.98,1.20, 1.38$. Three example single images $\text{OD}_i(x,y)$ are shown in the left column, plotted in the optical density (OD). The corresponding averaged profiles $\overline{\text{OD}}(x,y)$ are shown in the middle column. The right column shows the contrast between $\text{OD}_i(x,y)$ and $\overline{\text{OD}}(x,y)$, visualizing the emergence of the stripe-like density modulation in the roton-instable regime.}
    \label{highedd}
\end{figure*}

\section*{\bf Toward the strongly dipolar regime} Since the violation of $SO(2,1)$ symmetry arises from the non-local part of DDIs, the breathing mode frequency effectively characterizes the strongly dipolar regime where more significant violations occur when $\epsilon_{dd}>1$. To study the breathing mode in this regime, we tuned $\epsilon_{dd}$ by adjusting $a_s$ via Feshbach resonance. As shown in Fig.~\ref{highedd}a, an enhanced up-shift of the breathing mode frequency $\omega_B$ is observed with increasing $\epsilon_{dd}$. The degree of this enhancement depends on the dipole orientation and reaches a maximum at $\theta=90^{\circ}$, which can be explained by the corresponding reduction in the effective local interaction strength $\widetilde{g}_{\text{eff}}$, thus enhancing the role of the nonlocal DDIs.

Fig.~\ref{highedd}a summarizes the measured breathing oscillation frequencies alongside GPE and MCTDHB theoretical predictions. Discrepancies between GPE and MCTDHB become increasingly significant when approaching the unstable regime. Compared to the GPE results, which is plotted as solid lines, the MCTDHB analysis generally improves the quantitative accuracy by shifting the predicted frequencies upward. This correction is already noticeable at $\theta = 70^{\circ}$ for $\varepsilon_{dd} \geq 1.2$, and becomes essential to reconcile theory with experiment at $\theta = 90^{\circ}$. $\theta=90^\circ$ could be an ideal condition to further investigate the effect of quantum fluctuations in the 2D regime. As expected, the GPE simulations break down in the regime where both $\theta$ and $\epsilon_{dd}$ are high enough for the system becomes unstable without the beyond-mean-field contribution, which stabilizes the gas against the instability induced by attractive DDIs~\cite{chomaz2025quantum}. We measure $\omega_B$ of two samples within this mean-field unstable regime, with $\theta=70^{\circ}$ and $\epsilon_{dd}
=1.33, 1.38$. The results continue to show a deviation from the expected $SO(2,1)$ symmetry value. Notably, the MCTDHB simulation with $M = 2$ remains stable at higher values of $\epsilon_{dd}$ with respect to the GPE, providing a reliable theoretical estimate for the breathing mode frequency that aligns well with experimental observations. The breathing frequency of the sample with $\theta=80^{\circ}$ is slightly down-shifted from both the MCTDHB and GPE simulation, which we attribute to the  larger quench amplitude employed in the measurement~\cite{dalfovo1997frequency}.


At large $\varepsilon_{dd}$, stripes patterns also emerge in the experimental sample, as presented in Fig.~\ref{highedd}b. The density fluctuations become more pronounced with increasing $\epsilon_{dd}$ and manifests as a stripe-like modulation aligned with the dipole orientation within the 2D plane. In particular, $\widetilde{g}_{\text{eff}}$ remains small yet positive in this regime, indicating that the instability does not stem from a global collapse due to the attractive isotropic local interactions. Instead, this modulational instability is likely associated with the predicted emergence of a roton instability~\cite{mishra2016dipolar}. Such roton-like instability is a hallmark of the anisotropic nonlocality of DDIs in a strongly confined dipolar gas~\cite{santos2003roton}.

\section*{\bf Characterizing 2D strongly dipolar Bose gases} 
\begin{figure}
    \includegraphics[width=1.0\columnwidth]{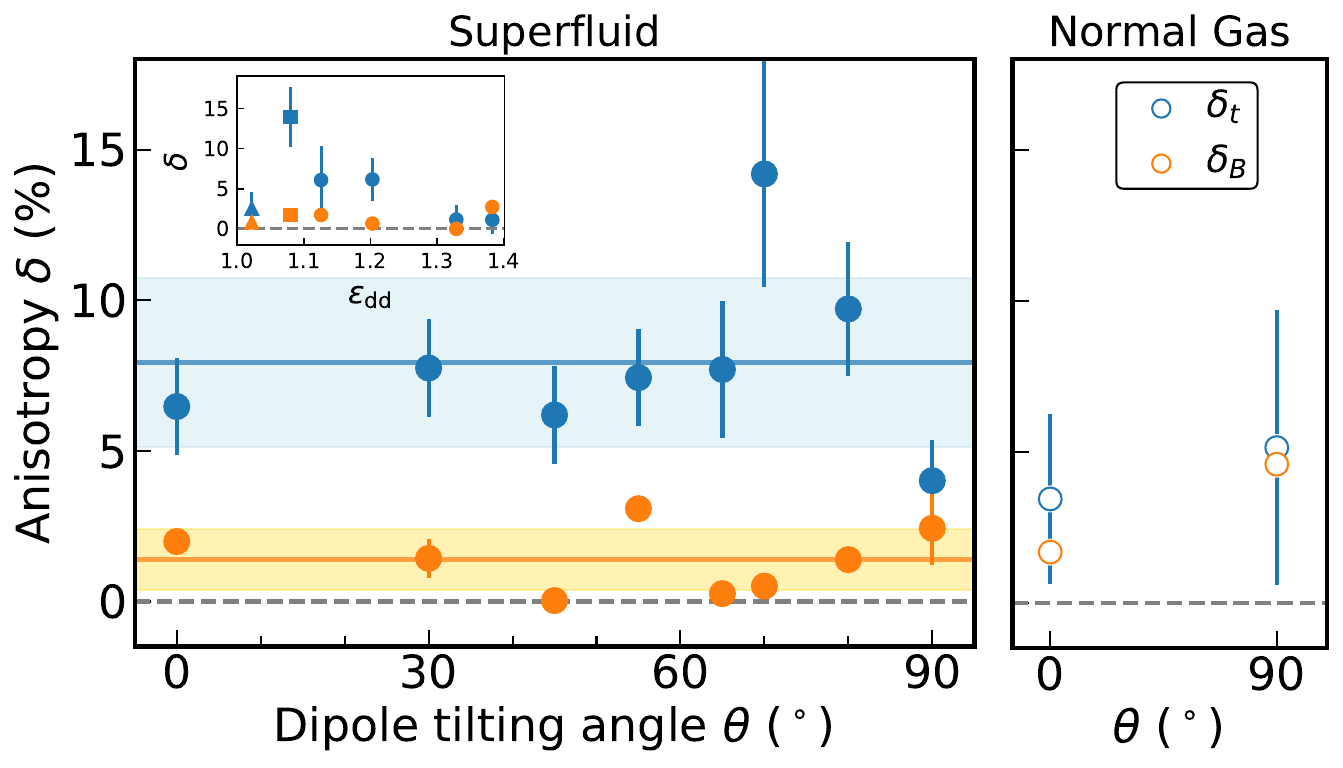}
    \caption{\footnotesize \textbf{Anisotropy of the breathing mode in the hydrodynamics and collisionless regime.} 
    The anisotropy of breathing mode $\delta_B$ and the trap frequency anisotropy $\delta_T$ of samples with $\epsilon_{dd}=0.98$ and different dipole angles are shown in orange and blue points. The left (right) panel demonstrates the condensate (normal gas) sample. The blue (orange) solid lines and shades represent the mean and the standard deviation of the trap (breathing) anisotropy. The dashed line is a guide to eyes. Error bars represent the 95$\%$ confidence interval. The inset shows the data of samples with $\epsilon_{dd}>1$ and dipole angles at 70$^{\circ}$ (circles), 80$^{\circ}$ (squares) and 90$^{\circ}$ (triangles).
    }
    \label{3Dfig}
\end{figure}

In previous work~\cite{he2025exploring}, quasi-long-range coherence associated with BKT superfluid was observed in a 2D dipolar system with moderate $\epsilon_{dd}\sim$0.5 and tilted dipole angles. Here, we demonstrate that the strongly dipolar sample exhibits hydrodynamic behavior at low temperature but becomes collisionless at higher temperature. We investigate the anisotropy of the breathing frequencies $\delta_B=|\omega_{bx}/\omega_{by}-1|$, where $\omega_{bx}$ and $\omega_{by}$ denote the oscillation frequencies of $\sigma_x$ and $\sigma_y$, respectively. In weakly interacting condensates or strongly interacting normal gases, the breathing mode exhibits the phase-locking oscillation, i.e., $\delta_B\approx 0$, as described by hydrodynamic theory~\cite{pethick2008bose,de2015collective}. This signature of hydrodynamic nature, previously observed in 2D Fermi gases~\cite{vogt2012scale,holten2018anomalous}, is also clearly present in our superfluid samples as shown in Fig.~\ref{3Dfig}. Despite significant trap anisotropy $\delta_T$, the measured $\delta_B$ in the condensate remains close to 0 and shows no correlation with $\delta_T$ or dipole orientation. These findings suggest the superfluid nature of our samples, even when $\widetilde{g}_{\text{eff}}$ is small at large $\theta$. In contrast, samples in the normal gas regime exhibit finite breathing anisotropy comparable to the trap anisotropy, consistent with decoupled oscillations along two directions in the collisionless limit~\cite{vogt2012scale}.    

\section*{\bf Outlook and Conclusion}
With tilting dipoles, we observe the breaking of scale invariance in coherent quasi-2D strongly dipolar gases, which is a result from the dominance of nonlocal interaction over local interaction. While the phase-locking behavior of the breathing mode reflects the superfluid hydrodynamics nature of the samples, the emergent stripe-type density modulations in the unstable regime indicate the onset of roton, possibly providing a precursor of a 2D stripe supersolid~\cite{aleksandrova2024density,SanchezBaena2025Tilted}.

We show that the explicit breaking of the $SO(2,1)$ symmetry by the nonlocal dipolar interaction term is closely tied to the emergence of quantum coherence and many-body correlations. In the thermal regime, the measured breathing mode frequency remains consistent with the scale-invariant prediction $\omega_B = 2\omega_t$ (see empty circles in Fig.~\ref{tiltfig}). However, significant deviations from this value consistently coincide with the onset of strong quantum fluctuations. This is evidenced by the growing discrepancy between the MCTDHB calculation and the GPE predictions. Notably, we are able to resolve this effect experimentally at $90^\circ$ (see Fig.~\ref{highedd}).

In this perspective, the breathing mode frequency measurement can be used to distinguish between different quantum-stabilized phases due to their distinctive physical properties~\cite{tanzi2019supersolid,chisholm2024probing}, thus paving the way for studying 2D supersolids in a weakly-interacting dilute dipolar gas. The breaking of scale invariance can also be of interest from the perspective of cosmology. As suggested in~\cite{cha2017probing,chandran2025expansion}, the scale invariance in a quasi-2D dipolar gas equipped with roton spectrum can validate different possible cosmology models in a scale that is unapproachable by observation, thus opening the possibility of using quasi-2D dipolar gases as the analog cosmology simulator~\cite{tian2018roton,tian2022probing}.

\clearpage
\newpage
\vspace{.1in} \noindent
\textbf{Methods}
\vspace{.05in} \\

{\bf Preparation of a strongly dipolar gas of $^{166}$Er in a quasi-2D trap} In this experiment, we prepare a quasi-2D sample of $^{166}$Er using a red-detuned laser sheet and an optical dipole trap~\cite{Seo2023,he2025exploring}. The sheet beam generates tight vertical confinement with a typical trap frequency of $\omega_z=2\pi \times 1100$ Hz.  With a typical atom number of $N\approx 2.8 \times 10^4$, the sample reaches a temperature of $T\approx 50$ nK. The chemical potential $\mu$ remains below 30 nK, depending on the dipolar angle, thereby satisfying the quasi-2D criterion $\mu < \hbar \omega_z \sim T$. The radial confinement in the x-y plane is provided by the optical dipole trap. We control the dipole orientation by applying an external magnetic field in the y-z plane. $\theta$ denotes the angle between the dipole orientation and the z-axis. 

\vspace{10pt}
{\bf Scale invariance in 2D} In the absence of the trap, conformal symmetry is a consequence of the scaling behavior of the mean-field interaction energy~\cite{saint2019dynamical},
\begin{equation}
    E_{\text{interaction}}=E_{\text{contact}}\propto \widetilde{g_{0}}\int n(r)^2dr^2,
\end{equation}
where $n(r)$ is the 2D density and $\widetilde{g_{0}}$ the dimensionless interaction strength $g_{0}$. Due
to its dimensionless nature $\widetilde{g_{0}}$ does not change by a spatial rescaling of the physical system. In other words, the action $\int dt \left\{ T+E_{\text{contact}}\right\}$, where $T$ is the total kinetic energy of the gas, remains invariant under the scaling transformation $r\to \lambda r, t\to\lambda^2t$. However, scale invariance only holds at the semiclassical level. Upon the inclusion of quantum fluctuations one realizes that a contact potential is ill-defined in 2D and a regularized interaction potential has to be introduced to describe the 2D gas~\cite{coon2002anomalies}.
The regularization naturally leads to the emergence of a bound state breaking scaling invariance~\cite{olshanii2010example}. This phenomenon, known as quantum anomaly, results in an upward shift of the breathing mode frequency with respect to the semiclassical value $2\omega_t$, where $\omega_t$ is the frequency of the trap. 

\vspace{10pt}
{\bf Breathing mode measurements} We excite the breathing mode by quenching the power of the radial trap, as illustrated in Fig.~\ref{fig1}c. The cloud is then allowed to evolve in the trap for varying hold times before being imaged along the z-axis. From the in-situ density profile, we characterize the breathing mode by the average cloud width $\sigma_{r}=\sqrt{\sigma_{x} \sigma_{y}}$, where $\sigma_x$ and $\sigma_y$ are the cloud widths along the two axes, obtained from a 2D Gaussian fit~\cite{supplementary}. We perform a damped sinusoidal fit to the oscillation of $\sigma_r$ to extract the breathing mode frequency $\omega_B$. The typical amplitude of the breathing mode ranges from 6$\%$ to 14$\%$ of the equilibrium cloud width. Each data point represents an average over 3-6 experimental realizations. Due to residual optical power from the sheet beam within the 2D plane, the trap exhibits a typical anisotropy of $\delta_T=|\omega_{y}/ \omega_{x}-1|=5\sim10\%$, which varies as a result of the trap drift. In the contact gas, this level of anisotropy is expected to induce a slight up-shift of the breathing frequency on the order of $0.7\%$ relative to $2\omega_{t}$~\cite{de2015collective, supplementary}.

\vspace{10pt}
{\bf MCTDHB calculations} Within the MCTDHB framework, the many-body wave function is represented as a superposition of permanents constructed from the non-interacting single-particle orbitals, whose amplitudes are variationally optimized to minimize the total energy~\cite{MCTDHB}. By increasing the number $M$ of orbitals, one can, in principle, describe quantum many-body systems in the strongly correlated regime. However, due to the high computational cost of large-$M$ calculations, typical analyses are restricted to $M\gtrsim 2$. We adopt the same approach, restricting our analysis to the cases $M=1$ (corresponding to the standard GPE) and $M=2$, which has already proven effective in capturing the dynamics of Bose gases with modulated interactions~\cite{nguyen2019parametric}. 
To account for anisotropic dipolar interactions in two dimensions, we have extended and modified the publicly available MCTDH-X code~\cite{Molignini2025MCTDHX,Lin2020QST,mctdhx-releases-2025} accordingly.
The short-range singularity of the anisotropic dipolar potential is addressed via a regularization scheme~\cite{supplementary}. Following the many-body simulations, breathing mode frequencies are extracted from the oscillation period of the collective kinetic energy.
\vspace{10pt}
\paragraph*{\bf Acknowledgement}
We acknowledge helpful discussions with Qi Zhou, Tin-Lun Ho and Chengdong He. This work has been supported by the RGC (RFS2122-6S04,C4050-23G). This research was also funded by the Swiss National Science Foundation (SNSF) grant numbers 200021 207537 and 200021 236722 and the Swiss State Secretariat for Education, Research and Innovation (SERI). GBJ acknowledges support from the Gordon and Betty Moore Foundation, grant DOI 10.37807/GBMF13794.


%

\end{document}


\preprint{APS/123-QED}

\title{Supplementary: Breaking of scale invariance in a strongly dipolar 2D Bose gas}

\author{Haoting Zhen}
\thanks{These authors contributed equally.}
\affiliation{Department of Physics, The Hong Kong University of Science and Technology,
Clear Water Bay, Kowloon, Hong Kong}%

\author{Yifei He}
\thanks{These authors contributed equally.}
\affiliation{Department of Physics, The Hong Kong University of Science and Technology,
Clear Water Bay, Kowloon, Hong Kong}%

\author{Sampriti Saha}
\affiliation{Institut f\"ur Theoretische Physik, ETH Z\"urich, Wolfgang-Pauli-Str. 27 Z\"urich, Switzerland.}%

\author{Mithilesh K. Parit}
\affiliation{Department of Physics, The Hong Kong University of Science and Technology,
Clear Water Bay, Kowloon, Hong Kong}%

\author{Mingchen Huang}
\affiliation{Department of Physics, The Hong Kong University of Science and Technology,
Clear Water Bay, Kowloon, Hong Kong}%

\author{Nicol\`o Defenu}
\affiliation{Institut f\"ur Theoretische Physik, ETH Z\"urich, Wolfgang-Pauli-Str. 27 Z\"urich, Switzerland.}%

\author{Gyu-Boong Jo}
\altaffiliation{email: gbjo@rice.edu}
\affiliation{Department of Physics and Astronomy, Rice University, Houston, TX, USA}%
\affiliation{Smalley-Curl Institute, Rice University, Houston, TX, USA}%
\affiliation{Department of Physics, The Hong Kong University of Science and Technology,
Clear Water Bay, Kowloon, Hong Kong}%

\newcommand{\ssaha}[1]{{\color{ForestGreen} #1}}
\date{\today}

\maketitle


\section{Experiment details}
We generate the quasi-2D dipolar gas using the same procedure described in our previous work~\cite{he2025exploring}. We utilize two image systems to characterize the sample. One is a low-resolution side-image system which images the sample after 16ms time-of-flight (TOF) on the x-z plane. We extract the atom number and temperature of the sample by performing a bimodal fitting on the TOF images. The second image system is built based on a NA=0.28 objective, which collects the insitu density of the sample on the x-y plane. We use this top-down high-resolution image system to measure the motion of the sample on the quasi-2D x-y plane from which we extract the frequencies of the dipole modes and the breathing mode. 

Besides estimating the 2D chemical potential by Eq \ref{chemical potential}, we can extract the chemical potential by fitting the insitu density profile~\cite{he2025exploring}. We fit the thermal tail of the azimuthal average profile of $\theta=0^{\circ}$ sample, which admits the least anisotropy generated by magnetostriction effect~\cite{chomaz2022dipolar}. The 2D chemical potential is below 30 nK, which qualitatively matches with the result in Fig \ref{crossover}.   

\section{Extracting the trap frequency}
\begin{figure}
    \centering
    \includegraphics[width=0.87\columnwidth]{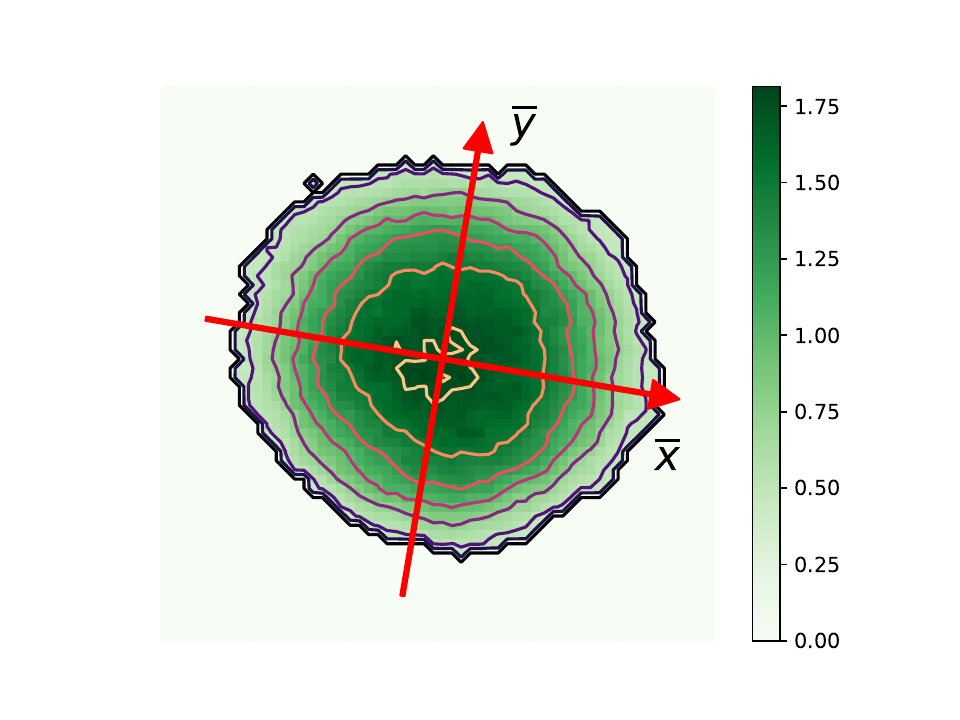}
    \caption{The trap principal axes extracted from the search. Two red arrows denote the direction of the trap principal axes extracted from the search. The background shows the averaged insitu density of $\theta=0^{\circ}$ sample. The contour lines are shown as a guide to the eye. The colorbar denotes the optical density.}
    \label{figaxes}
\end{figure}

We measure the trap frequency by extracting the frequency of the dipole mode excited by the quench of radial trap. The dipole mode can be characterized by the oscillatory motion of the cloud center after the quench. Since the trap principal axes do not align with the camera axes, the oscillation in the camera frame is a composition of the sinusoidal oscillations along two trap principal axes, which can be modeled by~\cite{fletcher2015bose},
\begin{equation}
    \binom{x}{y}=\begin{pmatrix}
\cos{\alpha} & -\sin{\alpha} \\
\sin{\alpha} & \cos{\alpha}
\end{pmatrix}\binom{\overline{x}}{\overline{y}}.
\end{equation}
$(x,y)$ and $(\overline{x},\overline{y})$ are coordinates in the camera frame and trap frame. $\alpha$ is the angle between the camera axes and trap principal axes. The frequency of the dipole modes extracted in the camera frame may systematically deviate from the real trap frequency. To find $\alpha$, we perform a search between $[-\pi/2,\pi/2]$ and determine $\alpha$ that maximizes the fitting quality $R^2$ of a undamped cosine fit. To validate this search, we compare the direction of the trap principal axes with the density profile of a static 2D sample with $\theta=0^{\circ}$, as shown in Fig \ref{figaxes}. 
In this case, the DDIs is isotropically repulsive in the 2D plane, which diminishes the magnetostriction effect that deviates the sample distribution from the trap geometry~\cite{chomaz2022dipolar}. The trap principal axes match well with the shape of the example profile, which proves the validity of the search for $\alpha$. From the oscillation frequencies of $\overline{x}$ and $\overline{y}$, $\omega_{dx}$ and $\omega_{dy}$, we extract the trap frequencies along principal axes as $\omega_{x,y}\equiv \omega_{dx,dy}$. The average trap frequency is computed as $\omega_t=\sqrt{\omega_x \omega_y}$. 

We employ a resampling method to evaluate the fitting uncertainty~\cite{kao2021topological}. This method randomly selects a portion of the data points for multiple times and performs a fitting on each selected subset. The uncertainty is represented by the standard deviation of these fitting results. In our work, we resample 80$\%$ of the dataset for 1000 times to generate the corresponding fitting uncertainty. Same analysis applies to the extraction of breathing frequency $\omega_B$, explained in the next section.

\begin{figure}
    \includegraphics[width=0.9\columnwidth]{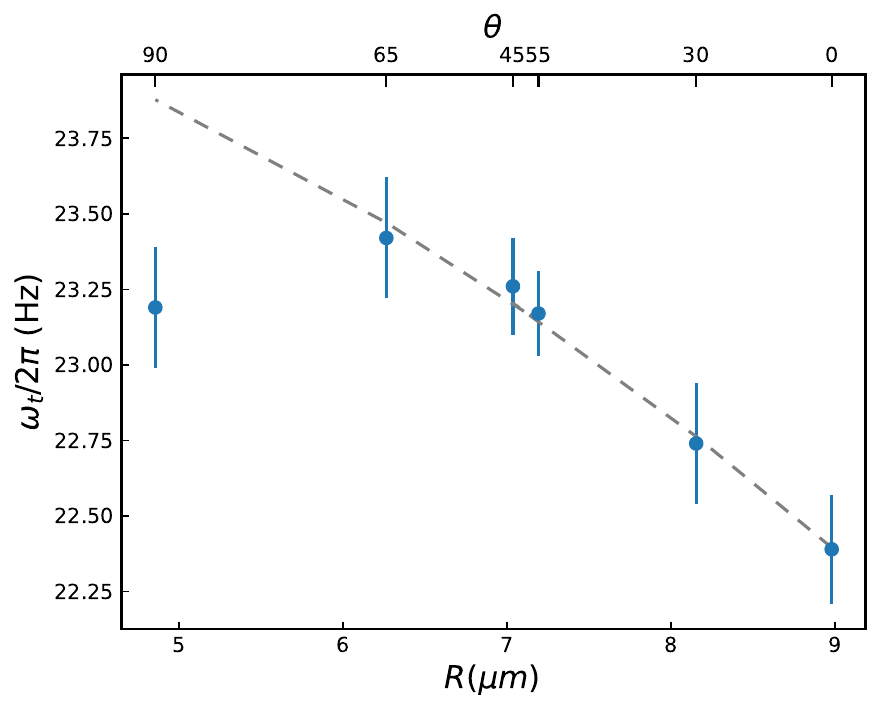}
    \caption{The trap frequency $\omega_t$ with different dipole orientation $\theta$ and sample Gaussian width $R$. The grey dashed line shows the fitting based on the anharmonicity estimate from the sample size~\cite{holten2018anomalous}. The error bars represent 2$\sigma$ fitting error.}
    \label{anitrapfreq}
\end{figure}

Due to the daily fluctuation of the optical traps, we observe a fluctuation in $\omega_t$ between different data, which does not affect the evaluation of the breathing ratio $\omega_B/\omega_t$ since we simultaneously extract $\omega_t$ and $\omega_B$ for each data point. For the data with similar trapping, we observe a systematic shift of $\omega_t$ with different dipole orientation $\theta$, which we attribute to the anharmonicity arise from the finite sample size~\cite{holten2018anomalous}. Specifically, if the sample size is smaller than the gaussian beam waist of the optical trap, the optimal trap frequency $\omega_t$ is expected to demonstrate a quadratic relationship with the sample size. In Fig \ref{anitrapfreq}, we plot $\omega_t$ with respect to the Gaussian width $R$ of the average profile of samples with different $\theta$. 
When $\theta$ increases, the effective local interaction decreases, which causes the shrink of sample size, resulting in an increasing $\omega_t$. The measured $\omega_t$ matches well with this quadratic model except the one with $\theta=90^{\circ}$. We attribute this deviation to the fluctuation of trapping.

\section{Trap Anisotropy}
Due to the residual radial potential of the tight-confining sheet beam, the radial trap is anisotropic. The trap frequency anisotropy $\delta_t=|\omega_y/\omega_x-1|$ is ranging from $3\sim14\%$, which varies due to the drift of optical potentials. This anisotropy of trap is expected to produce an effect on the breathing frequency, since it induces a coupling between the breathing mode and a quadrupole mode. In a 2D weakly-interacting contact gas, the breathing frequency is given by~\cite{de2015collective},
\begin{equation}
    \omega_B=\sqrt{\frac{3(\omega_x^2+\omega_y^2)}{2}+\frac{\sqrt{9(\omega_x^2+\omega_y^2)^2-32\omega_x^2\omega_y^2}}{2}}.
\end{equation}
In the isotropic case $\omega_x=\omega_y=\omega_t$, the ratio $\omega_B/\omega_t=2$ reflects the $SO(2,1)$ symmetry. With $\delta_t=0.1$, the ratio $\omega_B/\omega_t\approx2.014$ upshifts for around 0.7$\%$. We conclude that the upshift of breathing frequency from the trap anisotropy is small compared to the upshift caused by nonlocal DDIs.
\section{Characterizing the breathing behavior}
To characterize the breathing mode excited in the samples, we observe the oscillation of the sample size, represented by its average width $\sigma_r=\sqrt{\sigma_x\sigma_y}$. $\sigma_x$ and $\sigma_y$ is the sample width along X and Y axis, generated by performing a 2D Gaussian fitting on the sample profile at each hold time. 

We applied a damped cosine fitting to extract the breathing frequency $\omega_B$ from the oscillation of $\sigma_r$, which can be modeled by~\cite{kao2021topological},
\begin{equation}
    \sigma_r(t)=A_0e^{-k_1t}\cos{(\omega_B t+\phi)+k_2t+c},
\end{equation}
where $\phi$ is a constant phase, $k_1$ is the damping rate of the breathing mode and $k_2$ captures the change of sample size caused by heating or atom loss during the holding period. The relative breathing amplitude is defined as $A_0/c$. 

While the trap frequency $\omega_t$ has an average anisotropy $\delta_t$ of around 8$\%$, we observe small anisotropy in the breathing frequency along x and y axis, extracted from the oscillation of $\sigma_x$ and $\sigma_y$, which is defined as $\delta_B=|\omega_{b,y}/\omega_{b,x}-1|$ where $\omega_{b,x}$ and $\omega_{b,y}$ are the oscillation frequency of $\sigma_x$ and $\sigma_y$. This is consistent with the hydrodynamic property of condensate where the interaction between atoms result in a phase-lock between the compressional behavior along two axis~\cite{vogt2012scale, holten2018anomalous}.

\section{Details of the GPE simulation}
In the deep superfluid regime where the system can be describable by classical field theory, we examine the dipolar Gross-Pitaevskii equation (dGPE) with dipole-dipole interactions (DDI) in a quasi-2D harmonic trapping system at the mean-field level~\cite{cai2010mean}. We incorporate DDI into the momentum space and get
\begin{widetext} 
\begin{equation}
    i\hbar\partial_t\psi_{2D}(x,y,t)=\left\{-\frac{\hbar^2}{2m}\nabla^2+V(x,y)+\int\frac{d\textbf{k}}{(2\pi)^2} e^{ikr\cos\phi_k}\left[ g_{\text{eff}}+3g_{dd}F_d(\frac{kl_z}{\sqrt{2}})\right ]n(\textbf{k}) \right\}\psi_{2D}(x,y,t),
\end{equation}
where
\begin{equation}
     F_d(x)=x e^{x^2}\text{erfc}(x)(-\cos^2{\theta}+\sin^2{\theta}\cos^2{\phi_k}),
    \label{fd}
\end{equation}
\end{widetext} 
where $g_{\text{eff}}=\frac{4\pi\hbar^2}{m}[(a_s+a_{dd}(3\cos^2\theta-1)]$ is the local coupling strength, $g_{dd}=\frac{4\pi\hbar^2}{m}a_{dd}$ is the 2D dipolar coupling strength, $\text{erfc}$ is the complementary error function, $\phi_k$ is the angle between $k$ and the projection of dipole orientation on the 2D plane, and $V(x,y)$ is the external harmonic potential. To calculate the breathing frequency, we solve the time-dependent GPE with GPElab~\cite{antoine2014gpelab,antoine2015gpelab} after a quench of the radial trap frequency. To estimate the local interaction and nonlocal interaction in the condensates, we first solve the ground state static density profile through imaginary time propagation, then use the close form of nonlocal interaction given by ~\cite{cai2010mean} to calculate the nonlocal interaction energy. When the $g_\textbf{eff}$ becomes close to or below zero at large $\theta$ and $\epsilon_{dd}$, the system enters the mean-field unstable regime and dGPE fails to give stable oscillation behavior. 

\section{Breathing mode in the 2D-3D Crossover regime}
DDIs are not the only source that breaks the scale invariance in this experiment. In a quasi-2D quantum gas, the effect of a small portion of excited atoms along the tightly confined direction is known to decrease the breathing mode frequency from standard value~\cite{merloti2013breakdown,holten2018anomalous,peppler2018quantum}. To evaluate this influence of the third dimension, we vary the vertical confinement energy $\hbar \omega_z$ and measure $\omega_B$ of the samples with $\theta=0^{\circ}$ and $\epsilon_{dd}=0.98$. The 2D chemical potential $\mu_{\text{2D}}$ is estimated to quantify this effect~\cite{merloti2013two},
\begin{equation}
    \mu_{\text{2D}}=2\hbar \omega_t \sqrt{\frac{N_0a_{\text{eff}}}{\sqrt{2\pi}l_z}},
    \label{chemical potential}
\end{equation}
where $N_0$ is the condensate atom number. 

\begin{figure}
    \includegraphics[width=0.9\columnwidth]{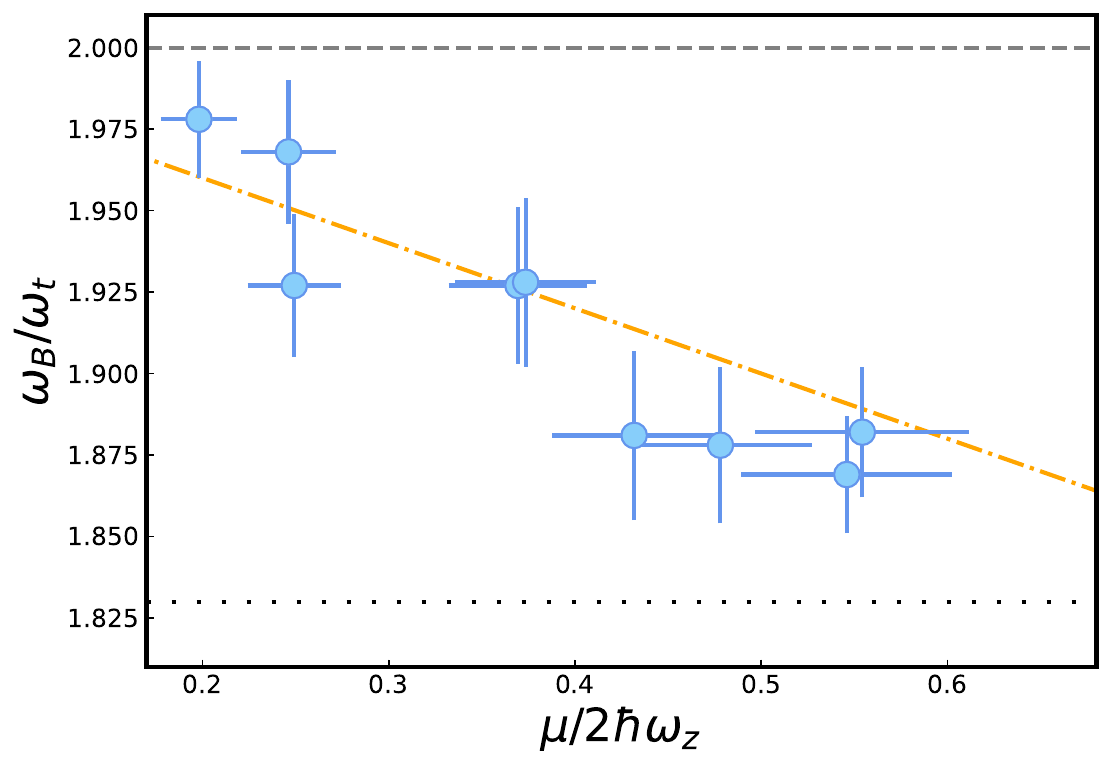}
    \caption{The breathing mode frequencies versus the
vertical confinement energy. The orange dashed line represents the
theoretical prediction~\cite{merloti2013breakdown}. The dashed and dotted lines represent
the prediction in the 2D and 3D limit~\cite{de2015collective}. The error bars represent 95$\%$ confident interval.}
    \label{crossover}
\end{figure}

As shown in Fig.~\ref{crossover}, our measurement qualitatively matches the theoretical model~\cite{merloti2013breakdown}. 
With the maximal trapping power, $\omega_B$ deviates from the 3D limit $\omega_B=1.83\omega_t$~\cite{de2015collective}, indicating the quasi-2D nature of our dipolar samples. Since $a_{\text{eff}}$ decreases with larger $\theta$, $\mu_{\text{2D}}$ decreases when dipoles rotate towards the 2D plane, which reduces the effect of the third dimension. We attribute the deviation from the theory to the finite temperature effect~\cite{de2015collective,holten2018anomalous} and the uncertainty of the $\mu_{\text{2D}}$ estimation. 
\section{Details of Simulations using MCTDH}
As the role of DDI grows in the experimental measurement a larger contribution of quantum fluctuations occurs, demonstrated by increasingly large deviation from the semiclassical GPE prediction. In order to partially describe the rise of strong quantum fluctuations, we implement the Multi-Configurational Time-Dependent Hartree method for Bosons (MCTDHB)~\cite{MCTDHB,MCTDHB2}, to solve the time-dependent (TDSE) and time-independent Schrödinger equations (TISE) for our quasi-2D dipolar system.

Instead of assuming condensation of all the bosons into a single orbital, in this framework, a multi-orbital configuration is considered. In MCTDHB, the ansatz expands the many-body state into a set of $M$ time-dependent single-particle orbitals ${\phi_j(\mathbf{r},t)}$ with occupation number ${n_j}$, such that $\sum_{j=1}^M n_j = N$. The many-body wavefunction can therefore be written as a superposition of time-dependent permanents~\cite{MCTDHB,MCTDHB2},
\begin{equation}
\label{eq:ansatz-general}
|\Psi(t)\rangle = \sum_{{n_j}} A_{n_1,\ldots,n_M}(t)|n_1,\ldots,n_M; t \rangle,
\end{equation}
where,
\begin{equation}
\begin{aligned}
|n_1,\ldots,n_M; t \rangle =
\frac{\prod_{j=1}^M \big(\hat{a}_j^\dagger(t)\big)^{n_j}}{\sqrt{n_1!\cdots n_M!}}\,|0\rangle,\\\hat{a}_j^\dagger(t) = \int d\mathbf{r}\,
\phi_j(\mathbf{r},t)\,\hat{\Psi}^\dagger(\mathbf{r})
\end{aligned}
\end{equation}
where $\phi_j(\mathbf{r},t)$ are the dynamically optimized
single-particle orbitals. In the $M\to \infty$ limit, the method captures the exact dynamics of the system, since the non-interacting permanent basis is a complete basis. Of course, in any practical numerical computation the sum has to be truncated to a finite number of orbitals $M$. The time-dependent coefficients $A_{n_1,\ldots,n_M}(t)$ and the orbitals $\phi_j(\mathbf{r},t)$ are the variational parameters which are determined by exploiting the Dirac–Frenkel principle~\cite{MCTDHB,MCTDHB2,MCTDH}. 

By allowing population of multiple, dynamically optimized orbitals, the ansatz itself takes into account the correlations between orbitals and coherent superpositions between configurations, hence capturing many-body quantum effects like fragmentation, depletion, and entanglement, beyond the mean-field Gross–Pitaevskii description, which is recovered at $M=1$. 

MCTDHB simulations are carried out for a perfect 2D plane and ignore the tight transverse confinement $\omega_{z}$, ensuring that confounding effects from the finite depth of the cloud are not included. We measure distances in units of the 2D confinement $l_x=\sqrt{\frac{\hbar}{m\omega_x}}$, where $\omega_x$ is the trap frequency along the \(x\)-axis. The parameters are rescaled as, $\rho \to l_x \rho$, $ t\to t/\omega_x$. The many-body Hamiltonian in harmonic oscillator units can hence be expressed as,
\begin{equation}
\begin{aligned}
    \hat{H}=\left[-\frac{1}{2}\nabla_i^2+\frac{1}{2}\big(x_i^2+\frac{\omega_y^2}{\omega_x^2 } y_i^2\big)\right]
+\Big[g_{2D}~\delta^{(2)}(\boldsymbol{\rho}_i-\boldsymbol{\rho}_j)\\   +V_{2D}^{\mathrm{reg}}(\boldsymbol{\rho}_i-\boldsymbol{\rho}_j)\Big],
\end{aligned}
\end{equation}
with $\boldsymbol{\rho}=(x,y)$ denoting planar coordinates. In the quasi-2D limit the contact coupling takes the form $g_{2D}=\frac{4\pi a_s}{\sqrt{2\pi\gamma}l_x}$ (in our dimensionless harmonic-oscillator units), where $a_s$ is the $s$-wave scattering length and $\gamma=\omega_x/\omega_z$, ratio of the trap frequencies along $x$ and $z$ directions.

In order to project the dipolar interactions to the 2D plane, one has to first consider the 3D expression~\cite{Stuhler2005Dipole, SanchezBaena2025Tilted},
\begin{equation}
V_{dd}({\boldsymbol{r}}_1 - {\boldsymbol{r}}_2) = \frac{C_{dd}}{4\pi} \, \frac{1-3\cos^2\theta}{|{\boldsymbol{r}}_1 - {\boldsymbol{r}}_2|^3},
\end{equation}where \(C_{dd} = \mu_0 \mu^2\) signifies the dipolar strength and 
\(\theta\) is the relative angle between the two dipoles.
In the quasi-two-dimensional regime, harmonic confinement along $z$ is strong enough such that the motion freezes into the ground state $\Phi_0(z)$, making the wavefunction separable. Hence, the effective interaction can be obtained by integrating out the z direction~\cite{ticknor2010quasi2D},
\begin{equation}
V_{2D}({\rho}) = \int dz \, dz' \, |\Phi_0(z)|^2 \, V_{dd}({\rho},z-z') \, |\Phi_0(z')|^2.
\end{equation}
To handle the short-distance divergence of the DDI in real space, we introduce a short-range cut-off. The regularised potential can be expressed as follows:
\begin{equation}
  V_{2D}^{\mathrm{reg}}(\mathbf{\rho})
  \;=\;
  \begin{cases}
    0, & \rho < \frac{\ell_z}{2},\\[4pt]
    V_{2D}(\mathbf{\rho}), & \rho \ge \frac{\ell_z}{2}
  \end{cases}
  \qquad (\rho = |\mathbf{\rho}|).
\end{equation}
Here, $\ell_z$ is the oscillator length of the tight $z$-confinement.
This cutoff helps us to overcome the unphysical singularity while maintaining the long-range anisotropic dipolar tail, which is responsible for the collective physics and low-energy scattering properties of the potential.

In our simulations, we stick to 2 orbitals, since, the occupation of first excited state is very low ($n_1/n_0 \ll 1$), we do not expect substantial corrections on addition of further orbitals. 
$M=2$ allows us to capture the leading quantum correlations, without being computationally expensive~\cite{nguyen2019parametric,Menotti2001DynamicSplitting}. The ansatz then reduces to
\begin{equation}
\label{eq:ansatz-2orb}
|\Psi(t)\rangle = \sum_{n=0}^{N} A_{n}(t)
\frac{(\hat{a}_1^\dagger(t))^n(\hat{a}_2^\dagger(t))^{N-n}}{\sqrt{n!(N-n)!}}|0\rangle,
\end{equation}
leading to coupled equations of motion for ${A_n(t)}$ and for the orbitals ${\phi_1,\phi_2}$. 
We set the atom number to $N=2000$ and re-scale the scattering length and dipolar lengths accordingly. Unlike GPE, for $M>1$ the number of particles affect the numerical stability of the MCTDHB~\cite{orbitals}. In practice, $N\sim 10^3$ ensures stable convergence of two-orbital ($M=2$) calculations.

We obtain the many-body ground state by evolving Eq.~\eqref{eq:ansatz-2orb} (orbitals and coefficients) in imaginary time $t=i\tau$. We claim the results have converged, and hence ground state has been obtained, when the total energy and the natural-orbital occupations become stationary within the defined tolerance range,
\begin{equation}
\bigl|E^{(k)}-E^{(k-1)}\bigr|<\delta\varepsilon_E,\qquad
\bigl|n_\alpha^{(k)}-n_\alpha^{(k-1)}\bigr|<\delta\varepsilon_n,
\end{equation}
where \(n_\alpha\) denotes the occupation of the \(\alpha\)-th natural orbital, and
$k$ denotes the index of the 
$k$-th time step (or iteration). 
In our computations we use $\delta\varepsilon_n = 10^{-12}$ and $\delta\varepsilon_E = 10^{-8}$. Employing the ground state, obtained via imaginary time propagation, as the initial condition, we then implement real-time propagation of the coupled MCTDHB equations to study the breathing dynamics. 

The non-linear coupled equations for the orbitals are integrated using the Runge–Kutta method of order 8. At the same time, for the coefficient part, the linear coupled equations are integrated by implementing the Davidson diagonalization for imaginary time evolution, while the MCTDH short-iterative Lanczos (MCS) propagator is used for real time propagation. Spatial discretization is performed with a discrete variable representation (DVR).
The \texttt{MCTDH-X} package~\cite{Molignini2025MCTDHX,Lin2020QST} is used for both ground-state relaxations and real-time evolutions. We have extended and modified the available \texttt{MCTDH-X} code~\cite{mctdhx-releases-2025}  to incorporate the regularized anisotropic dipolar interactions in two dimensions, which enables us to perform the simulation of quantum many-body effects in these systems.

To extract the radial breathing frequency, we monitor the oscillation of kinetic energy after the quench and determine its period. 
We then normalize this frequency by the geometric average of the radial trapping frequency 
\( \omega_R = \sqrt{\omega_x \omega_y} \). 
The resulting frequencies, together with experimental data, are presented in the main text; within uncertainties, the theory shows good agreement across the parameter range.


%